\documentclass[%
reprint,
superscriptaddress,
amsmath,amssymb,
aps,
amsthm,
pra,
]{revtex4-1}

\usepackage{graphicx}
\usepackage{dcolumn}
\usepackage{bm}
\usepackage{braket}%
\usepackage{theorem}
\usepackage{multirow}
\usepackage{enumerate}
\usepackage{framed}
\usepackage{bbm}

\usepackage{color}

\definecolor{memo}{RGB}{128,0,255}
\definecolor{gray}{RGB}{128,128,128}

\theorembodyfont{\upshape}

\renewcommand{\hat}{}
\newcommand{\hA}{\hat{A}}
\newcommand{\hB}{\hat{B}}
\newcommand{\hC}{\hat{C}}

\newcommand{\hE}{\hat{E}}

\newcommand{\hV}{\hat{V}}

\newcommand{\hX}{\hat{X}}

\newcommand{\hZ}{\hat{Z}}

\newcommand{\hPi}{\hat{\Pi}}
\newcommand{\hPhi}{\hat{\Phi}}

\newcommand{\hrho}{\hat{\rho}}

\newcommand{\mC}{\mathcal{C}}

\newcommand{\mH}{\mathcal{H}}

\newcommand{\mQ}{\mathcal{Q}}
\newcommand{\mR}{\mathcal{R}}
\newcommand{\mS}{\mathcal{S}}
\newcommand{\mT}{\mathcal{T}}

\newcommand{\mX}{\mathcal{X}}

\newcommand{\mZ}{\mathcal{Z}}

\newcommand{\trho}{\tilde{\rho}}

\newcommand{\PS}{P_{\rm S}}

\newcommand{\ident}{\hat{1}}

\newcommand{\Real}{\mathbf{R}}

\newcommand{\QED}{\hspace*{0pt}\hfill $\blacksquare$}

\newcommand{\argmax}{\mathop{\rm argmax}}

\newcommand{\Tr}{{\rm Tr}}

\newcommand{\Ker}{{\rm Ker}}

\def\gauss_sym#1{{\lfloor #1 \rfloor}}

\renewcommand{\ident}{\mathbbm{1}}
\newcommand{\identA}{\ident_{\rm A}}

\newcommand{\w}{{(\omega)}}
\newcommand{\hBw}{\hB^\w}
\newcommand{\hPhiw}{\hPhi^\w}

\newcommand{\opt}{\star}
\newcommand{\PSO}{\PS^\opt}
\newcommand{\PSG}{\PS^{\rm MEM}}

\newcommand{\mHA}{\mH_{\rm A}}

\newcommand{\dB}{d_{\rm B}}

\newcommand{\TrB}{\Tr_{\rm B}}
\newcommand{\TrBC}{\Tr_{\rm BC}}

\newcommand{\mUQ}{\overline{\mQ}}
\newcommand{\mUQv}{\mUQ_{\rm v}}
\newcommand{\mLX}{\underline{\mX}}

\begin{document}

\preprint{APS/123-QED}

\title{Optimal Discrimination of Optical Coherent States Cannot
Always Be Realized by Interfering with Coherent Light, Photon Counting, and Feedback}

\affiliation{%
 Quantum Information Science Research Center, Quantum ICT Research Institute, Tamagawa University,
 Machida, Tokyo 194-8610, Japan
}%
\affiliation{%
 School of Information Science and Technology,
 Aichi Prefectural University,
 Nagakute, Aichi 480-1198, Japan
}%

\author{Kenji Nakahira}
\affiliation{%
 Quantum Information Science Research Center, Quantum ICT Research Institute, Tamagawa University,
 Machida, Tokyo 194-8610, Japan
}%

\author{Kentaro Kato}
\affiliation{%
 Quantum Information Science Research Center, Quantum ICT Research Institute, Tamagawa University,
 Machida, Tokyo 194-8610, Japan
}%

\author{Tsuyoshi \surname{Sasaki Usuda}}
\affiliation{%
 School of Information Science and Technology,
 Aichi Prefectural University,
 Nagakute, Aichi 480-1198, Japan
}%
\affiliation{%
 Quantum Information Science Research Center, Quantum ICT Research Institute, Tamagawa University,
 Machida, Tokyo 194-8610, Japan
}%

\date{\today}

\begin{abstract}
 It is well known that a minimum error quantum measurement for arbitrary binary optical coherent states
 can be realized by a receiver that comprises
 interfering with a coherent reference light, photon counting, and feedback control.
 We show that, for ternary optical coherent states,
 a minimum error measurement cannot always be realized by such a receiver.
 The problem of finding an upper bound on the maximum success probability
 of such a receiver can be formulated as a convex programming.
 We derive its dual problem and numerically find the upper bound.
 At least for ternary phase-shift keyed coherent states,
 this bound does not reach that of a minimum error measurement.
\end{abstract}

\maketitle


Optical state discrimination is
one of the most fundamental issues in quantum optics
and quantum information science.
Since coherent beams of laser light are commonly used for 
optical communication and sensing applications,
distinguishing optical coherent states as accurately as possible is an important task.
A quantum measurement that maximizes the success probability for coherent states
can be analytically or numerically derived.
However, it is a highly difficult problem how to physically implement such a measurement.

In 1973, Dolinar \cite{Dol-1973} proposed a receiver based on a combination of a beam combiner,
a local coherent light source, a photon detector, and a feedback circuit,
and showed that this receiver realizes a measurement, called a minimum error measurement (MEM),
that maximizes the success probability for binary coherent states.
This was later demonstrated experimentally \cite{Coo-Mar-Ger-2007}.
Following Dolinar's work, several theoretical and experimental attempts have been made
to realize a receiver distinguishing binary coherent states
\cite{Ger-2004,Tak-Sas-2008,Wit-And-Tak-Syc-Leu-2010,Tsu-Fuk-Fuj-Ino-2011,Ass-Poz-Pie-2011,Syc-Leu-2016,Ros-Mar-Gio-2016}.
Also, many receivers comprising interfering with a coherent reference light, photon counting,
and feedback or feedforward control, which we call Dolinar-like receivers,
have been proposed to distinguish more than two coherent states
\cite{Dol-1982,Yam-1991,Bon-1993,Guh-Hab-Tak-2011,Izu-Tak-Fuj-Dal-2012,Li-Zuo-Zhu-2013,Nai-Guh-Tan-2014},
and related experimental demonstrations have been reported
\cite{Mul-Usu-Wit-Tak-Mar-And-Leu-2012,Che-Hab-Dut-Laz-2012,Bec-Fan-Bau-Gol-2013}.
However, it has been a long-standing question whether a Dolinar-like receiver
can realize an MEM for more than two coherent states.
It should be mentioned that a more complicated receiver realizing an MEM
for more than two coherent states was proposed \cite{Sil-Guh-Dut-2013},
but this receiver requires a special-purpose quantum computer,
making it impractical at present.

\begin{figure}[tb]
 \centering
 \includegraphics[scale=0.8]{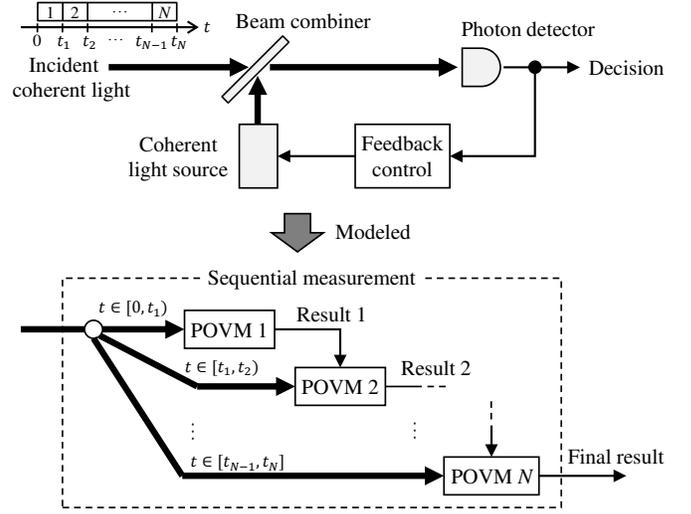}
 \caption{Dolinar-like receiver, which comprises interfering with a coherent reference light,
 photon counting, and feedback (or feedforward) control.}
 \label{fig:sequential}
\end{figure}

A coherent state $\ket{\alpha}$ of duration $T$ can be divided into $N$ time intervals
of duration $T/N$:
$\ket{\alpha} = \ket{\alpha/\sqrt{N}} \otimes \cdots \otimes \ket{\alpha/\sqrt{N}}$.
Let us consider a measurement, called a sequential measurement, on the $N$ systems
that is realized by carrying out local measurements on the individual systems adaptively,
where one adapts subsequent measurements based on the results of the previous ones.
A Dolinar-like receiver can be thought of as a sequential measurement
with the limit of $N \to \infty$ (see Fig.~\ref{fig:sequential}).
If $N$ divides $N'$, a sequential measurement on $N'$ systems is a special case of
that on $N$ systems.
In particular, a Dolinar-like receiver is a special case of a sequential measurement on any finite $N$ systems.
Thus, the maximum success probability of a Dolinar-like receiver
is upper bounded by that of such a sequential measurement.

In this paper, we investigate the maximum success probability
of a sequential measurement on two parties (i.e., $N = 2$), Alice and Bob.
As described above, this probability is an upper bound on that of a Dolinar-like receiver.
We show that the problem of obtaining this probability
can be reduced to an optimization problem with only Alice's measurement,
and that its dual problem can be easily derived.
An upper bound on the maximum success probability
of a sequential measurement for ternary phase-shift keyed (PSK) coherent states
is numerically computed using the dual problem.
We find that this upper bound is smaller than the success probability
of an MEM, which was obtained in Refs.~\cite{Cha-Ben-Hel-1989,Kat-Osa-Sas-Hir-1999}.
This means that, in the case of PSK coherent states,
a Dolinar-like receiver cannot realize an MEM,
which partially answers the above-mentioned long standing question.
This upper bound also tells us at least how large the difference between
the success probabilities of an optimal Dolinar-like receiver and an MEM.

To begin, we assume that Alice and Bob share a quantum system
that is prepared in one of $M$ known quantum states
given by density operators $\hrho_1, \cdots, \hrho_M$.
They try to distinguish them using the following sequential measurement.
Alice first performs a measurement, represented by
a positive operator valued measure (POVM) $\{ \hA_j \}$, on her system,
and sends the measurement result $j$ to Bob.
Then, Bob performs a measurement $\{ \hB^{(j)}_m \}$ on his system,
the choice of which depends on $j$.
The outcome $m \in \{ 1, \cdots, M \}$ of Bob's measurement
represents the final measurement result.
This sequential measurement is given by the POVM
$\{ \hPi_m = \sum_j \hA_j \otimes \hB^{(j)}_m \}$.
The conditional probability of obtaining the outcome $m$
given that the unknown state is $\hrho_k$ is $\Tr(\hrho_k \hPi_m)$.
Let $\xi_m$ be the prior probability for the state $\hrho_m$; then,
the success probability is $\PS \equiv \sum_m \xi_m \Tr(\hrho_m \hPi_m)$.
In order to maximize $\PS$,
we must optimize both Alice's and Bob's POVMs, i.e.,
$\{ \hA_j \}$ and $\{ \hB^{(j)}_m \}$.

In this paper, we recast this problem in the following way.
Each of Bob's POVM is uniquely labeled by an index $\omega$
\footnote{If Bob's quantum states span a $\dB$-dimensional space,
then each operator of Bob's POVM $\{ \hB_m \}$ can be represented by $\dB^2$ real numbers.
$\hB_M$ is uniquely determined by $\hB_1, \cdots, \hB_{M-1}$.
Thus, his POVM can be described by $(M-1)\dB^2$ real numbers,
which implies $\omega \in \Real^{(M-1)\dB^2}$.}.
Let $\{ \hBw_m \}$ be Bob's POVM indexed by $\omega$,
and $\Omega$ be the entire set of all possible values of $\omega$.
Alice first performs a continuous measurement $\{ \hA_\omega : \omega \in \Omega \}$
to determine which measurement Bob should perform,
and then sends the result $\omega$ to him.
He then performs the corresponding measurement $\{ \hBw_m \}$.
In this scenario, the sequential measurement is given by $\{ \hPi_m \}$ with
\begin{eqnarray}
 \hPi_m &=& \int_\Omega \hA(d\omega) \otimes \hBw_m. \nonumber
\end{eqnarray}
It is worth noting that any sequential measurement on two systems can be expressed
in this form.

Let us consider the problem of obtaining the maximum success probability
when only sequential measurements are allowed.
Since which measurement Bob performs is completely determined by the outcome $\omega$
of Alice's measurement $\hA$,
this problem can be formulated as the following optimization problem with only $\hA$:
\begin{eqnarray}
 \begin{array}{lll}
  {\rm P:} & {\rm maximize} & \displaystyle \PS(\hA) \equiv \sum_m
   \xi_m \Tr \left[ \hrho_m \int_\Omega \hA(d\omega) \otimes \hBw_m \right] \\
  & {\rm subject~to} & \hA \in \mR, \hA(\Omega) = \identA, \\
 \end{array} \nonumber
\end{eqnarray}
where $\mR$ is the entire set of $\hA$ satisfying
positivity (i.e., $\hA(\omega) \ge 0$ for any $\omega \in \Omega$)
and countable additivity (i.e., $\hA(\cup_k \omega_k) = \sum_k \hA(\omega_k)$ with mutually disjoint
$\{ \omega_k \} \subset \Omega$).
The above constraint, which states that $\hA$ must be a POVM, is convex,
and thus Problem~P is convex programming.
Let $\PSO$ be the optimal value of Problem~P.

According to the duality theory \cite{Boy-2009},
the dual problem of Problem~P provides an upper bound on $\PSO$.
To derive the dual problem, we construct the following Lagrangian function
\begin{eqnarray}
 L(\hA, \hX) &\equiv& \PS(\hA) + \Tr[ \hX [\identA - \hA(\Omega)]] \label{eq:L}
\end{eqnarray}
with $\hA \in \mR$ and $\hX \in \mS$,
where $\mS$ is the entire set of Hermitian operators.
In the case of $\hA(\Omega) \neq \identA$,
let $\hX = t \ket{x} \bra{x}$
with $\ket{x} \not\in \Ker[\identA - \hA(\Omega)]$;
then, $L(\hA, \hX)$ goes to $-\infty$ when
$t$ goes to $\infty$ or $-\infty$.
Thus, in this case, $\min_\hX L(\hA, \hX) = - \infty$.
This indicates
\begin{eqnarray}
 \max_{\hA} \min_\hX L(\hA,\hX) &=& \max_{\{\hA:\hA(\Omega)=\identA\}} \PS(\hA) = \PSO. \nonumber
\end{eqnarray}
Therefore, since $\max_x f(x,y) \ge \max_x \min_y f(x,y)$
always holds, we have
\begin{eqnarray}
 s(\hX) &\equiv& \max_{\hA} L(\hA,\hX) \ge \PSO. \label{eq:s_PS}
\end{eqnarray}
The dual problem is to minimize $s(\hX)$ over $\hX \in \mS$.
From Eq.~\eqref{eq:L}, $L(\hA, \hX)$ can be expressed as
\begin{eqnarray}
 L(\hA, \hX) &=& \Tr~\hX + \Tr \int_\Omega \left[ \TrB \sum_m \xi_m \hrho_m \hBw_m - \hX \right] \hA(d\omega).
  \nonumber
\end{eqnarray}
Now, let us introduce the following set:
\begin{eqnarray}
 \mX &\equiv& \left\{ \hX \in \mS : \hX \ge \TrB \sum_m \xi_m \hrho_m \hBw_m, ~ \forall \omega \in \Omega \right\}.
  \label{eq:mX}
\end{eqnarray}
In the case of $\hX \not\in \mX$, there exist
$\omega$ and a vector $\ket{x}$ satisfying
$\bra{x} (\TrB \sum_m \xi_m \hrho_m \hBw_m - \hX) \ket{x} > 0$.
In this case, letting $\hA(\omega) = t \ket{x} \bra{x}$
and taking $t$ to infinity yield $L(\hA, \hX) = \infty$;
i.e., $s(\hX) = \infty$.
In the other case (i.e., $\hX \in \mX$), $s(\hX) = \Tr~\hX$,
which is given by $\hA(\omega) = 0$ for any $\omega \in \Omega$.
Therefore, the dual problem can be rewritten as
\begin{eqnarray}
 \begin{array}{lll}
  {\rm DP:} & {\rm minimize} & \displaystyle \Tr~\hX \\
  & {\rm subject~to} & \hX \in \mX. \\
 \end{array} \nonumber
\end{eqnarray}
From Eq.~\eqref{eq:s_PS}, any feasible solution, $\hX \in \mX$, to Problem~DP
satisfies $\Tr~\hX \ge \PSO$.
By exploiting the convexity of Problem~P,
one can show that the gap between the optimal values of Problems~P
and DP is zero (see Appendix~\ref{append:gap}).
Note that, in the above discussion, we have considered Alice's measurement to be continuous,
of which a measurement with a finite number of outcomes is a special case.
But, we can see that there always exists an optimal sequential measurement
in which Alice performs a measurement with finite outcomes
(see Appendix~\ref{append:discrete}).

In order to show that, at least in some cases,
an MEM cannot be realized by any sequential measurement on two systems,
we will numerically show that the optimal value, $\PSO$, of Problem~P
is strictly smaller than the success probability (denoted as $\PSG$)
of the MEM.
To do this, it is sufficient to show $\Tr~\hX < \PSG$ for
a certain feasible solution $\hX \in \mX$ to Problem~DP.
However, whether $\hX \in \mX$ or not could be hard to say for a given $\hX$
in general, since $\mX$ is defined in terms of all Bob's POVMs.
Instead of $\mX$, we will use a subset $\mLX$ of $\mX$, as discussed later,
such that we can investigate whether $\hX \in \mLX$ in feasible computation.
Let us consider the following optimization problem:
\begin{eqnarray}
 \begin{array}{lll}
  {\rm DP':} & {\rm minimize} & \displaystyle \Tr~\hX \\
  & {\rm subject~to} & \hX \in \mLX. \\
 \end{array} \nonumber
\end{eqnarray}
Since $\mLX$ is a subset of $\mX$,
the optimal value of Problem~${\rm DP'}$ is not smaller than that of Problem~DP;
thus, any feasible solution $\hX \in \mLX$ to Problem~${\rm DP'}$
satisfies $\Tr~\hX \ge \PSO$.
We will compute the optimal value of Problem~${\rm DP'}$
as an upper bound of $\PSO$, and show that this value is smaller than $\PSG$.

We now examine the case of 3-PSK optical coherent states
$\{ \ket{\alpha_m} \}$ with equal probabilities,
where $\alpha_m = \alpha \exp(i2\pi m/3)$.
Let us divide the time duration of the input light into two equal time intervals;
the coherent state $\ket{\alpha_m}$ can be expressed as
$\ket{\alpha_m} = \ket{\beta_m} \otimes \ket{\beta_m}$ with $\beta_m = \alpha_m / \sqrt{2}$.
Substituting this into Eq.~\eqref{eq:mX} gives $\mX = \mC(\mQ)$,
where
\begin{eqnarray}
 \mC(\mQ) &\equiv& \left\{ \hX \in \mS : \hX \ge \sum_m \frac{q_m}{3} \ket{\beta_m} \bra{\beta_m},
                    ~ \forall \{ q_m \} \in \mQ \right\}. \nonumber \\
 \label{eq:CQ}
\end{eqnarray}
$\mQ \in \Real^3$ is the entire set of collections $\{ q_1, q_2, q_3 \}$ of
the conditional success probabilities
associated with Bob's measurement $\{ \hBw_m \}$ for the quantum states $\{ \ket{\beta_m} \}$;
i.e.,
\begin{eqnarray}
 \mQ &\equiv& \left\{ \{ q_m = \braket{\beta_m | \hBw_m | \beta_m} \} : \omega \in \Omega \right\}.
  \nonumber
\end{eqnarray}
It is easily verified that $\mQ$ is convex.

$\mQ$ is defined in terms of all Bob's measurements.
Instead of $\mQ$, we use a polyhedron $\mUQ$ that is a superset of $\mQ$.
How to construct $\mUQ$ will be described below.
From Eq.~\eqref{eq:CQ} and $\mUQ \supset \mQ$, $\mC(\mUQ) \subset \mC(\mQ) = \mX$.
Let $\mLX \equiv \mC(\mUQv)$, where $\mUQv$ is the entire set of vertices of the polyhedron $\mUQ$.
We can easily verify that $\mLX = \mC(\mUQ)$, since $\mUQ$ is convex,
and thus $\mLX \subset \mX$.
Since the number of elements of $\mUQv$ is finite,
whether $\hX \in \mLX$ can be numerically determined.

$\mUQ$ is constructed in the following way.
We choose finite points from the extremal points $\{ q_m \}$ of $\mQ$
(satisfying $q_m > 0$ for any $m$),
and then compute the tangent plane to $\mQ$ at each chosen point.
The tangent planes make the polyhedron $\mUQ$
\footnote{An extremal point of $\mQ$
is a collection of the conditional success probabilities
$\{ q_m = \braket{\beta_m | \hBw_m | \beta_m} \}$,
where $\{ \hBw_m\}$ is an MEM for $\{ \ket{\beta_m} \}$ with certain prior probabilities $\{ p_m \}$.
Thus, $\{ p_m \}$ is the normal vector of the tangent plane at this point.
This implies that each $\{ p_m \}$ determines the corresponding extremal point
$\{ q_m \}$ and the tangent plane at $\{ q_m \}$.
By computing MEMs for various $\{ p_m \}$,
we can construct $\mUQ$.
Note that an upper bound $\{ \overline{q}_m \}$ on the conditional success probabilities
$\{ q_m \}$ of an MEM is obtained
by the dual problem of the problem of finding the MEM.
Instead of $\{ q_m \}$, using $\{ \overline{q}_m \}$ ensures $\mUQ \supset \mQ$.}.
As the number of chosen points increases, $\mUQ$ tends to converge to $\mQ$;
i.e., the optimal value of Problem~${\rm DP'}$ tends to converge to that of Problem~DP.

We can efficiently compute the optimal value of Problem~${\rm DP'}$
by exploiting the symmetry that the states $\{ \ket{\beta_m} \}$ have.
Indeed, there exists a diagonal three-dimensional matrix $\hX$, in a certain fixed basis,
that is an optimal solution to Problem~${\rm DP'}$ (see Appendix~\ref{append:symmetry}).
This indicates that $\hX$ can be represented by only three real numbers.
Moreover, Problem~${\rm DP'}$ is convex programming;
thus, we can relatively easily compute the optimal value.

\begin{figure}[tb]
 \centering
 \includegraphics[scale=0.6]{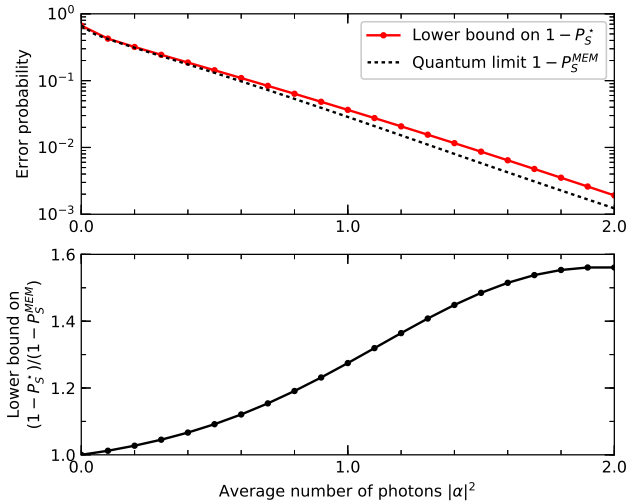}
 \caption{(a) A lower bound on the error probability of a sequential measurement
 for 3-PSK optical coherent states with equal prior probabilities.
 (b) The ratio of the lower bound to the quantum limit.}
 \label{fig:result_Pe}
\end{figure}

We have computed the optimal value of Problem~${\rm DP'}$ as
an upper bound on $\PSO$ using a polyhedron $\mUQ$ with about 100,000 vertices,
in the range of $|\alpha|^2 \le 2.0$,
where $|\alpha|^2$ is the average number of photons in the input light.
For visual convenience, instead of an upper bound on $\PSO$, we plot a lower bound on $1 - \PSO$,
i.e., the error probability of a sequential measurement.
The result is shown in Fig.~\ref{fig:result_Pe}.
We can see that this lower bound is larger than the error probability of an MEM
(called the quantum limit).
We remind that the error probability of a Dolinar-like receiver cannot be smaller than
this lower bound.
Therefore, this result concludes that, at least in this range,
any Dolinar-like receiver cannot realize an MEM.
Moreover, from this result, we cannot say that the difference between
the lower bound and the quantum limit is negligible;
in particular, in the range of $1.6 \le |\alpha|^2 \le 2.0$,
the lower bound is more than 1.5 times larger than the quantum limit.

We will now discuss multipartite systems.
These might be useful for computing a tighter bound,
since if $N > N'$, then the maximum success probability
of a sequential measurement on $N$ systems
(obtained by appropriately dividing the duration of the input light)
does not exceed that on $N'$ systems.
For simplicity, let us consider the tripartite case;
i.e., besides Alice and Bob, there is one more party, Charlie.
We consider a sequential measurement where
Alice, Bob, and Charlie perform measurements in this order.
Without loss of generality, a sequential measurement on Bob and Charlie
is given by a POVM $\{ \hPhi_m \}$ with
\begin{eqnarray}
 \hPhi_m &=& \sum_j \hB_j \otimes \hC^{(j)}_m, \nonumber
\end{eqnarray}
where $\{ \hB_j \}$ and $\{ \hC^{(j)}_m \}$ are respectively Bob's and Charlie's measurements
with finite outcomes.
Such a POVM can be uniquely identified by an index $\omega$,
as is in the bipartite case.
A sequential measurement on three parties is given by $\{ \hPi_m \}$ with
\begin{eqnarray}
 \hPi_m &=& \int_{\Omega_3} \hA(d\omega) \otimes \hPhiw_m, \nonumber
\end{eqnarray}
where $\Omega_3$ is the entire set of all possible values of $\omega$.
We can formulate the problem of obtaining the maximum success probability
by substituting $\hBw_m = \hPhiw_m$ and $\Omega = \Omega_3$ into Problem~P.
We can derive in the same manner as described above that
its dual problem is expressed as
\begin{eqnarray}
 \begin{array}{ll}
  {\rm minimize} & \displaystyle \Tr~\hX \\
  {\rm subject~to} & \hX \in \mX_3, \\
 \end{array} \nonumber
\end{eqnarray}
where
\begin{eqnarray}
 \mX_3 &\equiv& \left\{ \hX \in \mS : \hX \ge \TrBC \sum_m \xi_m \hrho_m \hPhiw_m,
                 ~ \forall \omega \in \Omega_3 \right\}. \nonumber
\end{eqnarray}
However, since $\mX_3$ is defined in terms of all sequential measurements
on Bob and Charlie,
computing the optimal (or near-optimal) value of the above dual problem is harder than
in the bipartite case.
A detailed investigation of multipartite systems is left for future studies.

Our technique of investigating the maximum success probability
of a sequential measurement can be generalized in several ways.
Obviously, this can be generalized to arbitrary prior probabilities.
Another generalization we can make is the case of several different states,
such as amplitude-shift keyed states or pulse-position modulated states.
By analyzing these states, we expect to be able to address the question of
which type of modulation is more effective when only sequential measurement
strategies are allowed.
Most of the ideas we proposed in this paper are applicable in these general settings.
Finally, generalization to other optimization criteria,
such as the Bayes criterion, the Neyman-Pearson criterion,
and their unambiguous (i.e., error-free) version, can be considered.
These topics are discussed in another publication
\cite{Nak-Kat-Usu-2017-seq_gen}.
Note that some results related to ours were independently
obtained by Croke {\it et al.} \cite{Cro-Bar-Wei-2017},
who gave a necessary and sufficient condition that a sequential
measurement maximizing the success probability must satisfy.

In summary,
we have derived the dual problem to the problem of
finding the maximum success probability of a sequential measurement,
and proposed a method of numerically computing
an upper bound on this probability by exploiting the dual problem.
We have also shown in numerical experiment that
an MEM of 3-PSK optical coherent states
cannot be realized by any sequential measurement in certain cases.
This indicates that a Dolinar-like receiver,
which comprises interfering with a coherent reference light, photon counting,
and feedback or feedforward control, could not realize an MEM.

We are grateful to O. Hirota of Tamagawa University for support.
T. S. U. was supported (in part) by JSPS KAKENHI (Grant No.16H04367).

\appendix

\section{Proof of zero duality gap} \label{append:gap}

We prove that the optimal values of Problems~P and DP are equal.
Let $\trho_m = \xi_m \hrho_m$.
When $\hX \in \mX$, $\Tr~\hX = s(\hX) \ge \PSO$ from Eq.~\eqref{eq:s_PS}.
Thus, it is sufficient to show that there exists $\hX \in \mX$ satisfying $\Tr~\hX \le \PS^\opt$.

Let us consider the following set:
\begin{eqnarray}
 \mZ &\equiv& \left\{ \left( \hA(\Omega) - \identA, u - \PS(\hA) \right) : (\hA, u) \in \mT \right\},
  \nonumber
\end{eqnarray}
where
\begin{eqnarray}
 \mT &\equiv& \left\{ \left( \hA, u \right) : \hA \in \mR, \PS^\opt < u \in \Real \right\}.
  \nonumber
\end{eqnarray}
Since $\PS(\hA) \le \PS^\opt < u$ when $\hA(\Omega) = \identA$,
$(0, 0) \not\in \mZ$.
Also, $\mZ$ is convex; i.e.,
if $(\hC_1, x_1), (\hC_2, x_2) \in \mZ$, then
$(t_1 \hC_1 + t_2 \hC_2, t_1 x_1 + t_2 x_2) \in \mZ$
for any $t_1, t_2 \ge 0$ with $t_1 + t_2 = 1$.
Indeed, let $(\hA_n, u_n)$ be $(\hA, u) \in \mT$
corresponding to $(\hC_n, x_n) \in \mZ$ for each $n \in \{ 1, 2 \}$
Also, let $\hA' = t_1 \hA_1 + t_2 \hA_2$ and $u' = t_1 u_1 + t_2 u_2$.
Then, $(\hA', u') \in \mT$ obviously holds, which gives
\begin{eqnarray}
 \lefteqn{ (t_1 \hC_1 + t_2 \hC_2, t_1 x_1 + t_2 x_2) } \nonumber \\
  &=& (\hA'(\Omega) - \identA, u' - \PS(\hA')) \in \mZ. \nonumber
\end{eqnarray}

Since $\mZ$ is a convex set with $(0, 0) \not\in \mZ$,
from separating hyperplane theorem (e.g., \cite{Boy-2009}),
there exists $(0, 0) \neq (\hZ, \alpha) \in \mS \otimes \Real$
such that $\Tr(\hZ\hC) + \alpha x \ge 0$ for any $(\hC, x) \in \mZ$.
Thus,
\begin{eqnarray}
 \Tr[\hZ [\hA(\Omega) - \identA]] + \alpha [u - \PS(\hA)] \ge 0 \label{eq:DPL_separation0}
\end{eqnarray}
for any $(\hA, u) \in \mT$.
Taking the limit $u \to \infty$ in Eq.~\eqref{eq:DPL_separation0} yields $\alpha \ge 0$.
We can show $\alpha > 0$ by contradiction.
We assume $\alpha = 0$.
From Eq.~\eqref{eq:DPL_separation0}, $\Tr[\hZ [\hA(\Omega) - \identA]] \ge 0$ holds for any $\hA \in \mR$,
which gives $\hZ = 0$.
This contradicts $(\hZ, \alpha) \neq (0, 0)$.

Let $\hX \equiv \hZ / \alpha$.
To complete the proof,
it is sufficient to show that such $\hX$ satisfies $\Tr~\hX \le \PSO$ and $\hX \in \mX$.
From Eq.~\eqref{eq:DPL_separation0}, we have
\begin{eqnarray}
 \Tr[\hX [\hA(\Omega) - \identA]] + u - \PS(\hA) \ge 0. \label{eq:DPL_separation}
\end{eqnarray}
Substituting $\hA(\omega) = 0$ $~(\forall \omega \in \Omega)$ into Eq.~\eqref{eq:DPL_separation} and
taking the limit $u \to \PS^\opt$ give $\Tr~\hX \le \PS^\opt$.
In contrast, let $\ket{a} \in \mHA$ and $t \in \Real$,
where $\mHA$ is Alice's Hilbert space.
Substituting $\hA(\omega) = t\ket{a}\bra{a}$ for certain $\omega \in \Omega$
and $\hA(\omega') = 0$ for any $\omega' \in \Omega$ with $\omega' \cap \omega = 0$
into Eq.~\eqref{eq:DPL_separation}
and taking the limit $t \to \infty$ give
\begin{eqnarray}
 \braket{a | \left[ \hX - \TrB \sum_m \trho_m \hBw_m \right] | a} &\ge& 0. \nonumber
\end{eqnarray}
Since this holds for any $\ket{a} \in \mHA$,
$\hX \ge \TrB \sum_m \trho_m \hBw_m$; i.e., $\hX \in \mX$.
\QED

\section{Alice's measurement with finite outcomes} \label{append:discrete}

We show that there exists an optimal solution to Problem~P
in which Alice performs a measurement with a finite number of outcomes.
Let $\hA^\opt$ be an optimal solution to Problem~P.
By using the results of Ref.~\cite{Chi-Dar-Sch-2007},
$\hA^\opt$ can be expressed by
\begin{eqnarray}
 \hA^\opt(\omega) &=& \int \hE^{(x)}(\omega) p(dx), \nonumber
\end{eqnarray}
where $p(x)$ is a certain probability density with a random number $x$
and $E^{(x)}$ is a POVM with finite support.
Let $x^\opt$ be a value satisfying
\begin{eqnarray}
 x^\opt &\in& \argmax_x \PS[\hE^{(x)}]. \nonumber
\end{eqnarray}
Then, we have
\begin{eqnarray}
 \PSO &=& \PS(\hA^\opt) \nonumber \\
 &=& \sum_m \xi_m \Tr \left[ \hrho_m \int_\Omega \int \hE^{(x)}(d\omega) p(dx) \otimes \hBw_m \right]
  \nonumber \\
 &=& \int \PS[\hE^{(x)}] p(dx) \le \PS[\hE^{(x^\opt)}]. \nonumber
\end{eqnarray}
Since $\PS[\hE^{(x^\opt)}] \le \PSO$ must hold,
$\PS[\hE^{(x^\opt)}] = \PSO$.
Thus, $\hE^{(x^\opt)}$ is an optimal solution to Problem~P
in which Alice performs a measurement with finite outcomes.
\QED

\section{Proof of existence of a symmetric optimal solution} \label{append:symmetry}

Suppose that we obtain $\mUQv$ such that,
for any $\{ q_m \} \in \mUQv$, each permutation of $\{ q_m \}$ (e.g., $\{ q_1, q_3, q_2 \}$) is
also in $\mUQv$.
Such $\mUQv$ can be easily obtained.
We show that, in a certain fixed basis,
there exists a three-dimensional diagonal matrix $\hX$ that
is an optimal solution to Problem~${\rm DP'}$.
The 3-PSK coherent states $\{ \ket{\beta_m} \}$ has a $Z_3$ symmetry; i.e.,
there exists a unitary operator $\hV$ with $\hV^3 = \identA$
such that $\ket{\beta_m} = \hV^{m-1} \ket{\beta_1}$.
Alice's Hilbert space $\mHA$ is chosen as the three-dimensional Hilbert space spanned by
the states $\{ \ket{\beta_m} \}$.
Here, we take the basis of eigenvectors of $\hV$.
Obviously, $\hV$ is a three-dimensional diagonal matrix.

Let $\hX^\opt$ be a three-dimensional matrix that is an optimal solution to Problem~${\rm DP'}$,
but not necessarily diagonal.
We have that for any $\{ q_m \} \in \mUQv$ and $k \in \{0, 1 ,2\}$,
\begin{eqnarray}
 \lefteqn{ \hV^k \hX^\opt \hV^{\dagger k} - \sum_m \frac{q_m}{3} \ket{\beta_m} \bra{\beta_m} } \nonumber \\
 &=& \hV^k \left[ \hX^\opt - \sum_m \frac{q_m}{3} \hV^{\dagger k} \ket{\beta_m} \bra{\beta_m} \hV^k \right]
  \hV^{\dagger k} \nonumber \\
 &=& \hV^k \left[ \hX^\opt - \sum_{m'} \frac{q'_{m'}}{3} \ket{\beta_{m'}} \bra{\beta_{m'}} \right] \hV^{\dagger k}
  \ge 0, \label{eq:VXV}
\end{eqnarray}
where $m' = m - k$ if $m > k$; otherwise, $m' = m - k + 3$.
$\{ q'_m \}$ is the permutation of $\{ q_m \}$ such that $q'_{m'} = q_m$.
The inequality follows from $\hX^\opt \in \mLX = \mC(\mUQv)$ and Eq.~\eqref{eq:CQ}.
It follows that $\hV^k \hX^\opt \hV^{\dagger k} \in \mLX$ from Eq.~\eqref{eq:VXV}.
Let $\hX \equiv \sum_{k=0}^2 \hV^k \hX^\opt \hV^{\dagger k} / 3$;
then, we can easily see that $\hX \in \mLX$.
Also, we have
\begin{eqnarray}
 \Tr~\hX &=& \frac{1}{3} \sum_k \Tr(\hV^k \hX^\opt \hV^{\dagger k})
  = \frac{1}{3} \sum_k \Tr~\hX^\opt = \Tr~\hX^\opt. \nonumber
\end{eqnarray}
Therefore, $\hX$ is also an optimal solution to Problem~${\rm DP'}$.
In contrast, since
\begin{eqnarray}
 \hV \hX \hV^\dagger &=& \frac{1}{3} \sum_k \hV^{k+1} \hX^\opt \hV^{\dagger k+1}
  = \hX, \nonumber
\end{eqnarray}
$\hX$ commutes with $\hV$; i.e., $\hX$ is diagonal.
\QED

%

\end{document}